\documentclass{emulateapj}

\usepackage{amsmath}
\usepackage{amssymb}
\usepackage{graphicx}
\usepackage{grffile}
\usepackage[usenames,dvipsnames]{color}
\usepackage{bm}
\usepackage{natbib}
\usepackage{hyperref}
\usepackage{tabularx}
\usepackage{xspace} 

\usepackage{array}
\newcolumntype{L}{>{\raggedright\let\newline\\\arraybackslash\hspace{0pt}}X}

\usepackage[english]{babel}

\newcommand{\mrm}[1]{\mathrm{#1}}
\newcommand{\pow}[1]{\ifmmode{}^{#1}\else ${}^{#1}$\fi}
\newcommand{\HI}{{\text{H\MakeUppercase{\romannumeral 1}}}\xspace}

\newcommand{\Lya}{\ifmmode{\mathrm{Ly}\alpha}\else Ly$\alpha$\xspace\fi}

\newcommand{\cm}{\,\ifmmode{{\rm cm}}\else cm\fi}

\newcommand{\ergps}{\,{\rm erg}\,{\rm s}\ifmmode{}^{-1}\else ${}^{-1}$\fi}
\newcommand{\Mpch}{\,{\rm Mpc}\,\ifmmode h^{-1}\else $h^{-1}$\fi}
\newcommand{\snru}{\,\ifmmode{\mathrm{Myr}^{-1}}\else Myr${}^{-1}$\fi}
\newcommand{\kms}{\,\ifmmode{\mathrm{km}\,\mathrm{s}^{-1}}\else km\,s${}^{-1}$\fi}

\def\lsim{~\rlap{$<$}{\lower 1.0ex\hbox{$\sim$}}}

\def\gsim{~\rlap{$>$}{\lower 1.0ex\hbox{$\sim$}}}
\newcommand{\xxx}[1]{\textcolor{blue}{\textbf{[...]}}}

\newcommand{\fc}{\relax\ifmmode{f_{\mathrm{c}}}\else  $f_{\mathrm{c}}$\xspace\fi}
\newcommand{\fccrit}{\relax\ifmmode{f_{\mathrm{c,\,crit}}}\else  $f_{\mathrm{c,\,crit}}$\xspace\fi}
\newcommand{\cl}{{\mathrm{cl}}}

\slugcomment{Draft from \today}

\begin{document}
\title{From Mirrors to Windows: Lyman-Alpha Radiative Transfer in a Very Clumpy Medium}

\author{Max Gronke$^{1}$, Mark Dijkstra$^{1}$, Michael McCourt$^{2,3}$, S. Peng Oh$^{2}$}
\affil{$^{1}${Institute of Theoretical Astrophysics, University of Oslo, Postboks 1029 Blindern, 0315 Oslo, Norway;}} 
\affil{$^{2}${Department of Physics, University of California, Santa Barbara, CA 93106, USA;}}
\affil{$^{3}${Hubble fellow.}}

\email{maxbg@astro.uio.no}

\begin{abstract}
Lyman-Alpha (Ly$\alpha$) is the strongest emission line in the
Universe and is frequently used to detect and study the most
distant galaxies. Because Lya is a resonant line, photons typically scatter prior to escaping; this scattering process complicates the interpretation of Ly$\alpha$ spectra, but also encodes a wealth of information about the structure
and kinematics of neutral gas in the galaxy.  Modeling the Ly$\alpha$ line
therefore allows us to study tiny-scale features of the gas, even in
the most distant galaxies.  Curiously, observed Ly$\alpha$ spectra can be
modeled successfully with very simple, homogeneous geometries (such
as an expanding, spherical shell), whereas more realistic, multiphase
geometries often fail to reproduce the observed spectra.  This seems
paradoxical since the gas in galaxies is known to be multiphase.  In
this Letter, we show that spectra emerging from \textit{extremely}
clumpy geometries with many clouds along the line of sight converge to
the predictions from simplified, homogeneous models.  We suggest that
this resolves the apparent discrepancy, and may provide a way to study
the gas structure in galaxies on scales far smaller than can be probed
in either cosmological simulations or direct (i.e., spatially-resolved) observations.
\end{abstract}


\section{Introduction}
\label{sec:intro}
Lyman-Alpha (Ly$\alpha$) is the strongest emission line in the Universe and is frequently used to study the highest redshifts \citep[for reviews, see, e.g.,][]{Barnes2014,Dijkstra2014review,Hayes2015}. However, the resonant nature of the \Lya line complicates studies of galaxies as \Lya observables do not directly connect to intrinsic properties of the gas. These complications also carry great potential for the study of \Lya emitting systems: at every scattering event the frequency of the \Lya photon is shifted slightly depending on the temperature and velocity of the surrounding \HI. As a consequence, information about the density and velocity fields of the gas is encrypted in the observed \Lya spectrum.

Curiously, observed \Lya spectra can be modeled quite successfully with very simple, homogeneous geometries such as an expanding, spherical shell \citep[e.g.,][]{Verhamme2008b,Yang2015} wheres more complex, multiphase geometries have difficulties in reproducing, for instance, the pronounced double peaked \Lya spectra commonly observed. 
This holds true for both simplified models of multiphase structure and for more sophisticated models such as snapshots from cosmological simulations
(see \S~\ref{sec:implications} and \citealp{Gronke2016a,fog_paper} for a more detailed discussion).
This is surprising considering that the interstellar and circumgalactic medium are well-known to be multi-phase -- which suggests that `clumpy' geometries are closer to reality than the homogeneous ones. In any case, the mismatch motivates us to revisit of the subject of resonant radiative transfer (RT) in multiphase geometries.

Previous studies on this subject have focused on `moderately clumpy media', that is, on average a few clumps per sightline. However, observations of quasar absorption spectra suggest the presence of a large amount of tiny ($\ll 1\,$pc) clumps in a variety of astrophysical systems \citep[see][and references therein]{McCourt2016}. 
Similar conclusions hold for large-scale quasar outflows
\citep[e.\,g.][]{Finn2014}, and BAL regions around AGNs
\citep[e.\,g.][]{Bottorff2000}.
These observations translate to models which contain orders of magnitude more (much smaller) clumps per sightline than has previously been studied in the context of resonant line transfer. 

In this Letter, we show that \Lya photons behave fundamentally differently in an environment with such extreme clumping than in other multiphase geometries. This alleviates (and possibly eliminates) the tensions between observed and modelled \Lya spectra discussed above. 

\begin{figure*}
  \centering
  \plotone{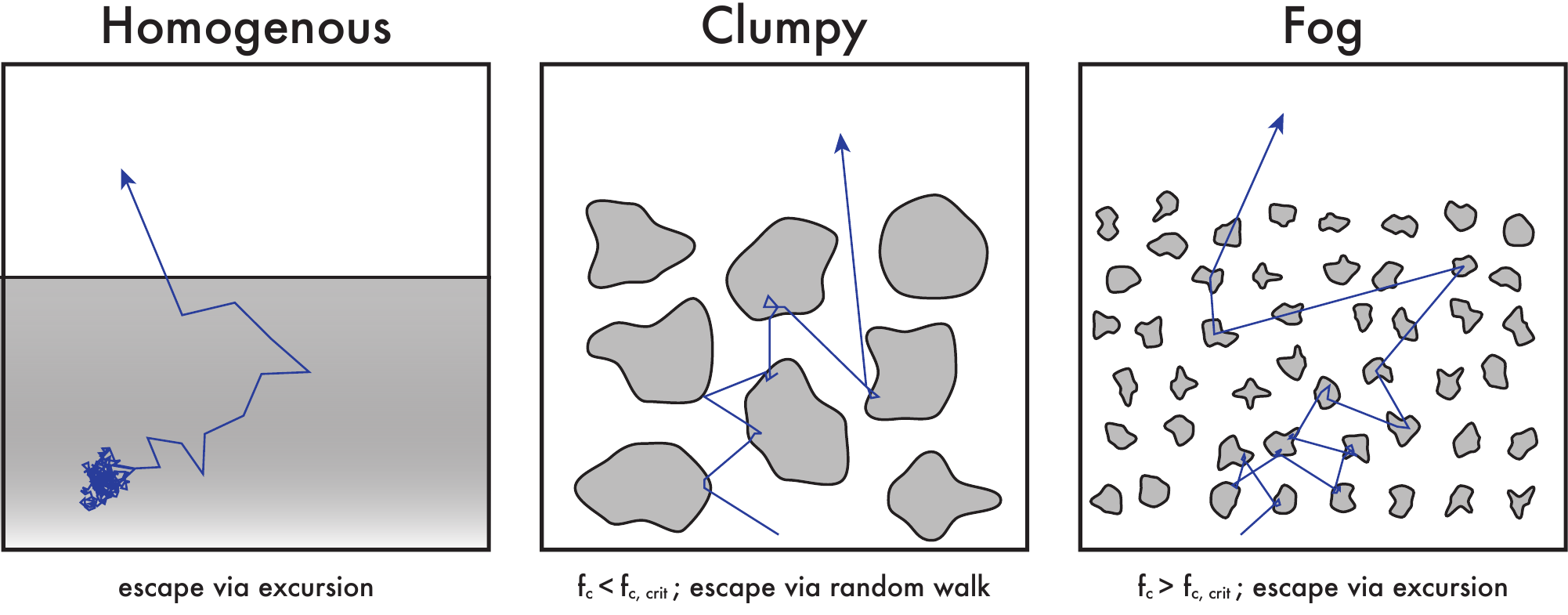}
  \caption{Sketch of the \Lya RT regimes discussed in this work.}
  \label{fig:sketch}
\end{figure*}

\section{The transition from clumpy to homogeneous behavior: Analytic insights}
\label{sec:analytical-estimates}
In this section, we first describe some basic concepts of \Lya RT in static homogeneous and clumpy media (in \S~\ref{sec:lya-rt_homogeneous_medium} and \S~\ref{sec:lya-rt_clumpy_medium}, respectively).
Then we show that under certain circumstances \Lya RT in a clumpy medium might behave the way it does in a homogeneous medium. And lastly (in \S~\ref{sec:bound-betw-clumpy}), we derive the conditions of this transition.
Fig.~\ref{fig:sketch} shows a visual summary of this section.

We express the frequency $x$ in units of Doppler shifts from line center (i.e., $x \equiv (\nu-\nu_0) / \Delta \nu$ where $\nu_0\approx 2.47\times 10^{15}\,\mathrm{s}^{-1}$ and $\Delta\nu\approx 1.06\times (T / 10^4\,\mathrm{K})^{1/2}$ are the line-center frequency and the thermal line width, respectively) and the Voigt parameter as $a_v = \Delta_{\mathrm{L}} / (2 \Delta\nu) \approx 4.70 \times 10^{-4} (T/10^{4}\,\mathrm{K})^{-1/2}$ in which $\Delta_{\mathrm{L}}\approx9.94\times 10^7\,\mathrm{s}^{-1}$.
Furthermore, we write the \HI scattering cross section simply as $\sigma_{\HI}(x, T) = \sigma_0(T) H(x)$ where $\sigma_0(T) \approx 5.90\times 10^{-14} (T / 10^4\,\mrm{K})^{-1/2}\cm^{2}$ and
$H(x)$ denotes the Voigt function which can be approximated in the wing of the line ($|x| > 3$) to be $H(x) \sim a_v / (\sqrt{\pi} x^2)$.
The geometry we consider is a semi-infinite slab of height $2 B$ and total line-center optical depth $\tau_0$ (between the half-height and surface of the slab).

\subsection{\Lya radiative transfer in a homogeneous medium}
\label{sec:lya-rt_homogeneous_medium}
\Lya RT in homogeneous medium has been studied for decades. The mean free path of Ly$\alpha$ photons at line center is very small, which forces photons to scatter frequently before escaping. Each scattering event changes the frequency of the photon due to random thermal motions of the hydrogen atoms. On rare occasions scattering occurs off a hydrogen atom far in the tail of the Maxwellian velocity distribution. 
A Ly$\alpha$ photon that gets scattered to $|x| \gsim 2.6$ can escape from a medium with $\tau_0 =10^3$ in a single flight. The time spent scattering is small and the escape time is of order the light crossing time \citep{adams75}: 
\begin{equation} 
t_{\rm exc} = \frac{B}{c}  
\label{eq:t_low_optical_depth} 
\end{equation}
where $c$ is the speed of light. 

For extremely opaque media, escape in single flight is not possible because the damping wings of the Voigt profile remain optically thick. We can estimate when this happens by recalling that boundary between the line core and wings is $x_{*} \approx 3.1$. Thus, the wings become optically thick when $\tau_{0} H(x_{*}) > 1$, i.e. $\tau_{0} a_{\rm v} > x_{*}^{2} \sqrt{\pi} = 17$, or $\tau_{0} > 4 \times 10^{4}$ for $T=10^{4}\,$K. 

Ly$\alpha$ scattering in the wing of the line profile ($|x| \gsim  3$) changes the frequency by an average amount $\langle \Delta x \rangle = -1/|x|$ with an r.m.s. $\sqrt{\langle \Delta x^2\rangle}=1$ \citep{Osterbrock1962}. For a photon at some frequency far in the wing, $x_{\rm w}$, it therefore takes $N_{\rm w} \sim x_{\rm w}^2$ scattering events to return to the line core, during which the photon traverses a distance $l\sim \sqrt{N_{\rm w}}\lambda_{\rm mfp}(x_{\rm w})$, where $\lambda_{\rm mfp}(x_{\rm w})$ denotes the mean free path at frequency $x_{\rm w}$. This sequence of wing scattering events is called an `excursion'. We can write $\lambda_{\rm mfp}(x_{\rm w})=\lambda_0/H(x_{\rm w})$, where $\lambda_0=B/\tau_0$ denotes the mean free path at line centre. A photon can escape when $l = B$, i.e. when $x_{\rm w}B/(\tau_0 H(x_{\rm w})) = B$, which yields a characteristic frequency for escape (Adams 1972, Harrington 1973, Neufeld 1990), $x_{\rm esc}$:  
\begin{equation}
x_{\mathrm{esc}} = (\tau_0 a_v / \sqrt{\pi})^{1/3}\;.
\label{eq:xesc}
\end{equation} At $|x|<x_{\rm esc}$ it is harder to escape in an excursion, while it is increasingly unlikely for photons to scatter to frequencies $|x| > x_{\rm esc}$. The spectrum of emerging photons therefore peaks at $x_{\rm esc}$. The escape time is:
\begin{equation}
t_{\mathrm{exc}} = \frac{N_{\rm w} \lambda_{\rm mfp} (x_{\mathrm{esc}})}{c} = \frac{B}{c} \left(\frac{\tau_0 a_v}{\sqrt{\pi}}\right)^{1/3}
= \frac{B x_{\rm esc}}{c}\;.
\label{eq:t_exc}
\end{equation}

The left panel of Fig.~\ref{fig:sketch} illustrates the escape path in a homogeneous medium. Initially, a large number of scattering events ($\sim \tau_0$) lead to practically no displacement from the emission site. Then, the photon's frequency shifts to $\sim x_{\mathrm{esc}}$ and it can escape easily (in $\sim x_{\mathrm{esc}}^2$ scatters).

\subsection{\Lya radiative transfer in a clumpy medium}
\label{sec:lya-rt_clumpy_medium}
For this case, we consider a slab filled with spherical clumps with radius $r_\cl$, gas temperature $T$ and neutral hydrogen column density $N_{\HI, \cl}$ (measured from the center to the boundary of the sphere). We characterize the number density of clumps in the slab by the covering factor \fc which denotes the number of clumps intercepted by a sightline orthogonal to the slab's surface between the half-height and the boundary of the slab. In this setup, such a sightline will intercept a column density of
\begin{equation}
N_{\HI, \mathrm{total}} = \frac{4}{3} \fc N_{\HI, \cl}
\label{eq:N_HI_total_clumps}
\end{equation}
where the $4/3$ is a geometrical factor.
Furthermore, we only consider the case of optically thick clumps without any `holes' in the setup ($\fc \gtrsim 1$). We will drop these constraints and consider the general case in \citet{fog_paper}.

Previous work showed that given this geometry, \Lya photons emitted (at line center) in the midplane of the slab escape by scattering off the clumps' surfaces and slowly diffusing outwards \citep{Neufeld1991,Hansen2005}. In this picture (sometimes referred to as `surface scatter approximation'), the clumps are optically thick to Ly$\alpha$ photons throughout the scattering process. Thus, the photons random walk between the clumps, which leads to $N_\cl \sim \fc^2$ clump interactions (shown in the central panel of Fig.~\ref{fig:sketch}). Here, the mean free path between clump interactions is simply the mean clump separation $B/\fc$. As a result, 
the escape time is given by
\begin{equation}
t_{\mathrm{rw}} = N_{\mathrm{sct, rw}} \frac{B}{c \fc} = \frac{B \fc}{c}\;.
\label{eq:t_rw}
\end{equation} 

Importantly, the trapping time depends on the number of scatters in the single longest excursion (\S\ref{sec:lya-rt_homogeneous_medium}) or between clouds (\S\ref{sec:lya-rt_clumpy_medium}), {\it not} the total number of scatters. For instance, for a very optically thick medium, \citet{Adams1972} found that it takes $\sim \tau_0$ core scattering events to be first scattered to $x_{\mathrm{esc}}$, something that has been reproduced with Monte-Carlo simulations \citep[see][]{Dijkstra2006ApJ...649...14D,Laursen2009ApJ...696..853L}. This is vastly larger than the number of scatters $\sim x_{\rm esc}^{2}$ we have invoked during an excursion. However, the mean free path associated with core scatters is small, $B/\tau_{0}$, so the time spent performing core scatters $\sim 1/c (B/\tau_{0})\tau_{0} \sim B/c$ is much less than the time $\sim x_{\rm esc}B/c$ spent on the excursion (Equation~\eqref{eq:t_exc}). Similarly, in a clumpy medium, the time spent on surface scatters is $\sim f_{\rm c}$ times longer than on core scattering. Thus, under certain conditions photons can scatter substantially within a single cloud $\mathcal{N}_{\rm scat} \sim 100-10^{4}$, but they still spend most of their time in a random walk in the inter-cloud medium.  

\subsection{The transition to a `very clumpy' medium}
\label{sec:bound-betw-clumpy}

Consider a photon emitted at line center. There are essentially two modes of escape. If it remains largely confined to the Doppler core $|x| < 3$, then it can escape by scattering off the clump surfaces, which act as mirrors, with an escape time given by Equation~\eqref{eq:t_rw}. Such photons have a peak at line center. On the other hand, as it scatters, the photon random walks in frequency as well. It can also potentially escape via its longest frequency excursion; the spectrum of such photons peak at $\pm x_{\rm esc}$. Both of these escape routes operate simultaneously. The corresponding relative flux of the central and off-center peaks in the spectrum is proportional to the relative escape rates, i.e. $F_{\rm rw}/F_{\rm exc} = t_{\rm exc}/t_{\rm rw}$. The flux ratios can also be interpreted as the relative probability that a photon will escape via one of these routes. The boundary between these regimes, when both have comparable fluxes, arises when $t_{\rm exc}=t_{\rm rw}$, or for a covering fraction: 
\begin{equation}
\fccrit =  \begin{cases}
\left(\frac{a_{v} \tau_{0}}{\sqrt{\pi}}\right)^{1/3} & \text{ for } a_v \tau_0 > 17\\
2.1 & \text{ otherwise.}
\end{cases}
\label{eq:fccrit} 
\end{equation} 
Here, the upper and lower expression is valid for the wing and core dominated regime, respectively\footnote{Due to the angular variation in pathlength we expect this expression to differ from numerical results by a factor of order unity \citep{adams75}.}.
The escape time and hence $f_{\rm c, crit}$ is independent of $\tau_{0}$ in the Doppler core dominated regime, when the photon escapes in a single flight. We evaluate the numerical value by demanding continuity at $a_{v} \tau_{0} = 17$. 

One can also understand this result by comparing the photon mean free paths $\lambda = B/\sqrt{N}$. Since $N=f_{\rm c}^{2}$ [$N=x_{\rm esc}^{2}$] for surface scatters [frequency excursion], $(\lambda_{\rm rw},\lambda_{\rm exc})=(B/f_{\rm c}, B/x_{\rm esc})$ for these two processes. Frequency excursions dominate the spatial diffusion of photons once $\lambda_{\rm exc} > \lambda_{\rm rw} = B/f_{\rm c}$, i.e. the mean free path during frequency excursions is larger than the mean spacing between clumps. This is only possible if the clumps are optically thin at $x_{\rm esc}$. From $\lambda_{\rm exc} > \lambda_{\rm rw} \Rightarrow x_{\rm esc} < f_{\rm c}$, this is indeed the case: 
\begin{equation} 
\tau_{\rm clump} = \frac{\tau_{0} H(x_{\rm esc})}{f_{\rm c}} < \frac{\tau_{\rm 0} a_{\rm v}}{\sqrt{\pi} x_{\rm esc}^{3}} = 1. 
\end{equation} 
This makes sense: once clumps become optically thin, then the discretization of cold gas into distinct clumps is immaterial: RT is wholly determined by the cumulative intercepted column density. As the dense clumps morph from reflecting mirrors to translucent windows, a continuum approximation becomes accurate, and the fog of cloudlets is indistinguishable from a continuous medium with the same optical depth profile. In other words, {\it \fccrit marks the transition where multiphase (clumpy) media affect Ly$\alpha$ photons as if the medium were perfectly homogeneous.} Importantly, numerically \fccrit is small enough that we expect to hit this limit frequently in astrophysically realistic environments. In the reminder of this {\it Letter}, we investigate the transition using Monte-Carlo RT simulations. 

\begin{figure}
  \centering
  \plotone{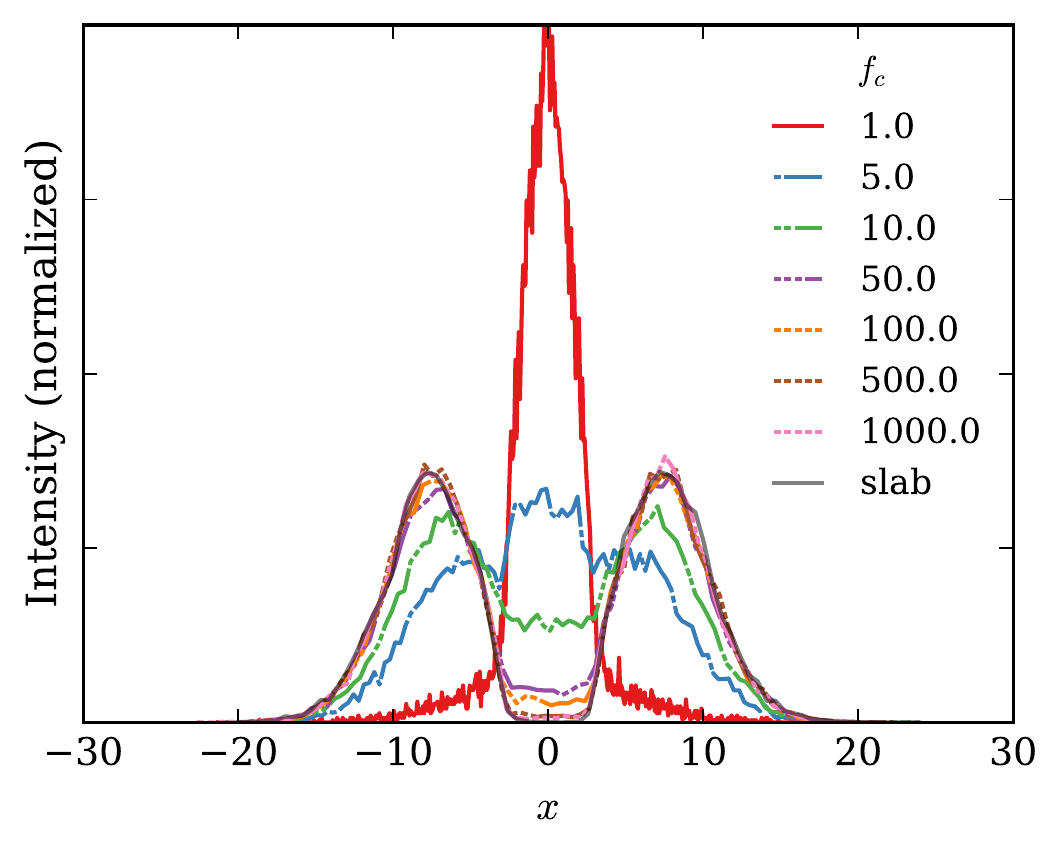}
  \caption{\Lya spectra for a constant total column density $N_{\rm HI,total}=4/3\times 10^{19}\,\cm^{-2}$ and various values of $\fc$. 
Increasing $\fc$ corresponds to a lower flux at $x=0$, and the grey line denotes the spectrum from a homogeneous slab with the same column density.\\}
  \label{fig:spectra_N_HI_total}
\end{figure}

\begin{figure}
  \centering
  \includegraphics[width=\linewidth]{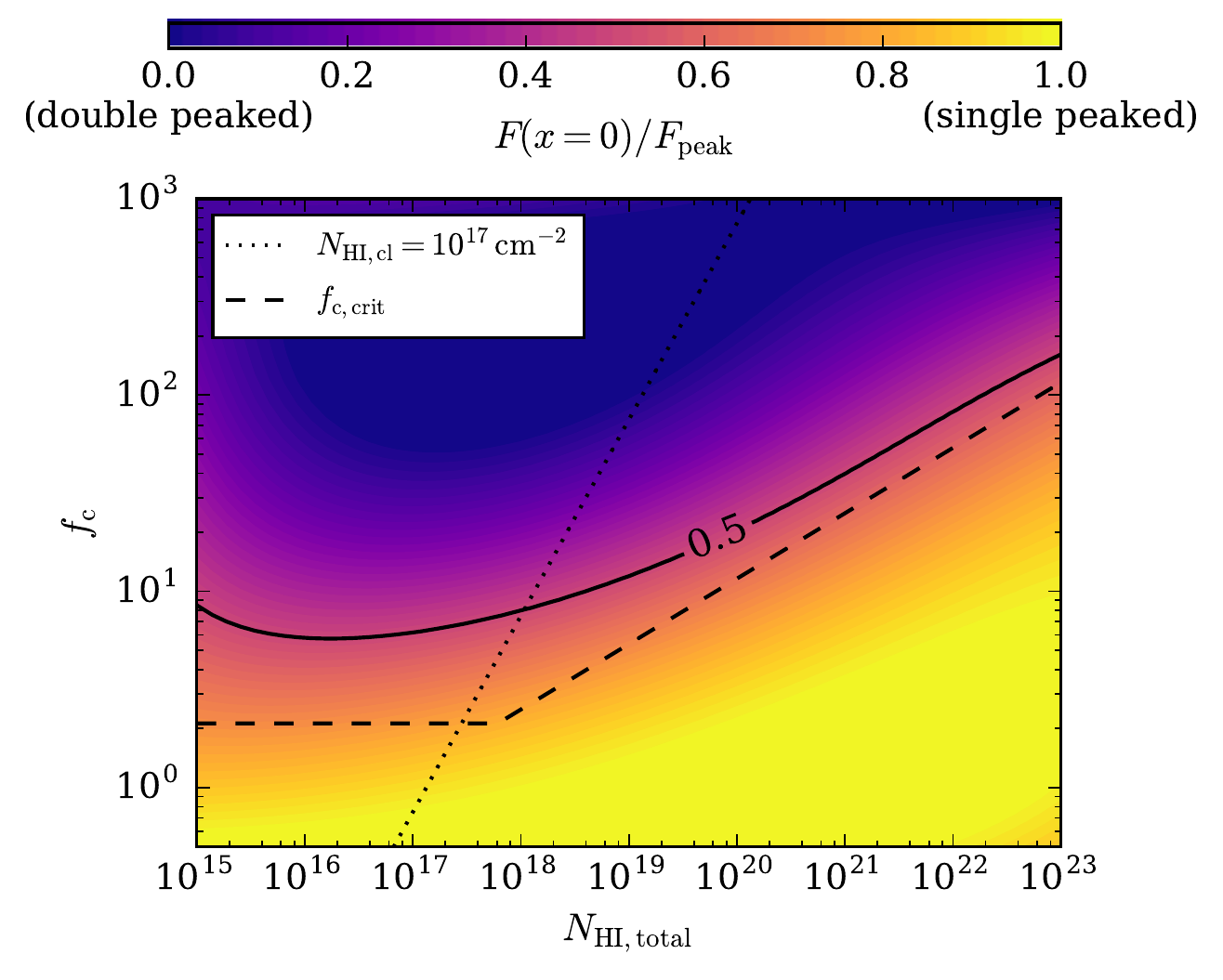}
  \caption{Overview of the transition in a clumpy medium. The $x$ and $y$ axis denote the total column density and the average number of clumps per sightline, respectively. The color coding shows the ratio between the flux at line center and the peak value. We also mark the $F(0) / F_{\mathrm{peak}} = 1/2$ value with a solid line and our theoretical estimate from Equation~\eqref{eq:fccrit} with a dashed line. The dotted line marks the clump column density as predicted by shattering \citep{McCourt2016}}.
  \label{fig:result_overview}
\end{figure}

\begin{figure*}
  \centering
  \plotone{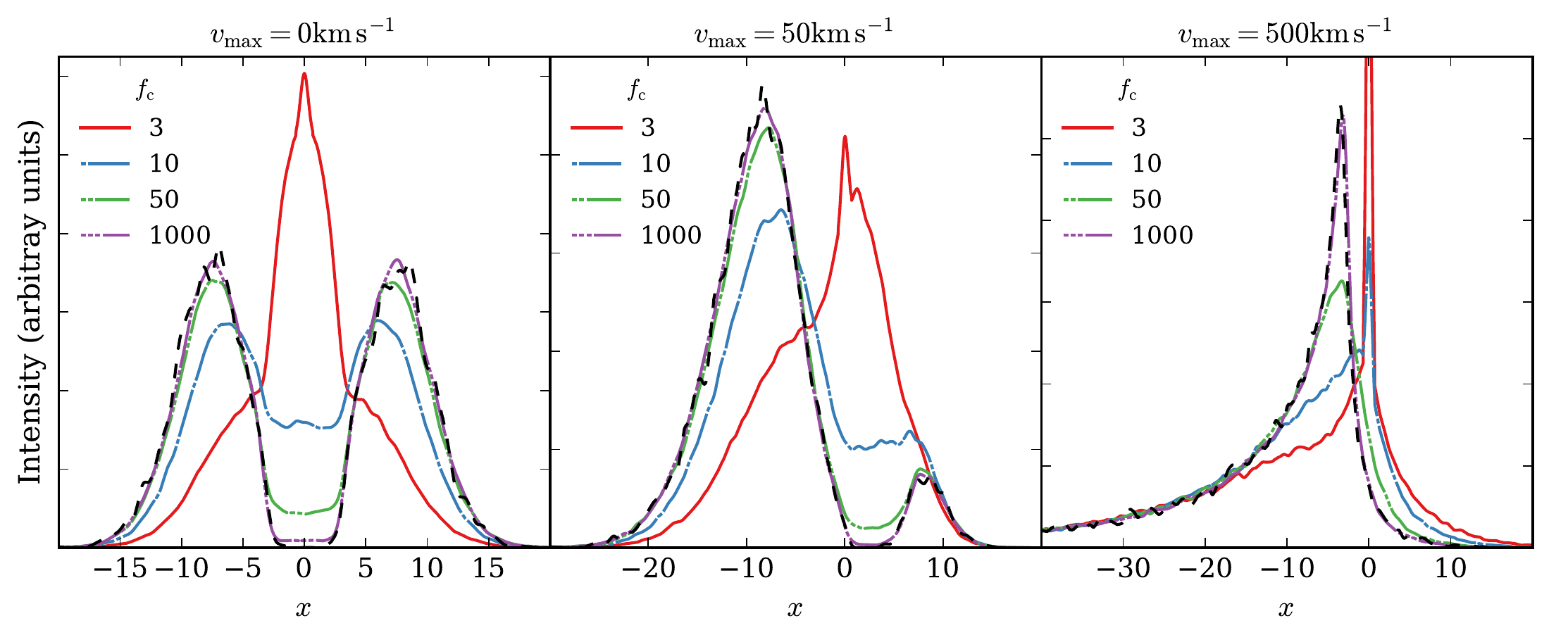}
  \caption{Spectra emergent from an outflowing slab with total column density $N_{\HI, \mathrm{total}}=4/3\times 10^{19}\cm^{-2}$. The panels show different outflow velocities (increasing from left to right), and in each panel the colored lines  show the spectra obtained from a clumpy medium with various covering factors. The black dashed line shows the spectrum from a homogeneous slab with matching total column density and outflow velocity.\\}
\vspace{.4em}
  \label{fig:outflows}
\end{figure*}

\section{Numerical Results}
\label{sec:results}
We solve the \Lya RT problem in clumpy media (discussed in Sec.~\ref{sec:analytical-estimates}) numerically using the Monte Carlo RT code \texttt{tlac} \citep{Gronke2014a}. 

We use the same setup as described in Sec.~\ref{sec:analytical-estimates} in which we fix $T=10^{4}\,$K, and $B=50\,$pc. 
The two main parameters that affect the RT are ({\it i}) the clump column density, $N_{\HI,\cl}$, which we vary from $10^{14}\cm^{-2}$ to $10^{22}\cm^{-2}$, and ({\it ii}) the covering factor, \fc, which we vary from $0$ to $1000$. In agreement with \citet{Hansen2005}, we find no separate further dependence on clump size once these parameters are specified. \\

Fig.~\ref{fig:spectra_N_HI_total} shows the \Lya spectra emerging from a clumpy medium with fixed total column density $N_{\HI, \mathrm{total}} = 4/3\times 10^{19}\cm^{-2}$, but for different covering factors \fc ranging from $1$ to $10^3$. One can clearly see the transformation of the spectrum with increasing \fc from a single peak \citep[with extended wings; see][for details]{Hansen2005}, to a double peak with relatively high flux at line center, to a double peak with no flux at line center. The latter is identical to the spectrum one obtains from a homogeneously filled slab with equivalent column density (shown in Fig.~\ref{fig:spectra_N_HI_total} as a {\it grey line}). 

The transition from a single to a double peaked spectrum shown in Fig.~\ref{fig:spectra_N_HI_total} can be quantified by the ratio of the flux at line center and at the peak, i.e., $F(x = 0) / F_{\mathrm{peak}}$. For low \fc, escape proceeds via surface scatterings, and the spectrum is single peaked (at line center). The ratio $F(x = 0) / F_{\mathrm{peak}}\sim 1$ for surface scattering. In contrast, for high \fc escape proceeds via damping wing excursions, which gives rise to classical double peaked spectra with low flux at line center.  The ratio $F(x = 0) / F_{\mathrm{peak}}\sim 0$ for escape in an excursion.\\

Fig.~\ref{fig:result_overview} shows the same flux ratio ($F(x = 0) / F_{\mathrm{peak}}$) through color coding for a range of different total \HI column densities $N_{\rm HI, \mathrm{total}}$, and covering factors \fc. Bright colors correspond to single peaked spectra, while dark colors correspond to pronounced double peaked spectra. One can read off from Fig.~\ref{fig:result_overview} that for a clump column density of $N_{\HI, \cl}\approx 10^{17}\cm^{-2}$ \citep[the value predicted by shattering; see][]{McCourt2016} the spectrum is single peaked below $\fc \sim 3$. For increasing \fc it transitions to a double peaked spectrum with low flux at line center at $\fc \gtrsim 10$. 

In addition, Fig.~\ref{fig:result_overview} also shows the analytical estimate from Equation~\eqref{eq:fccrit} for the transition value \fccrit (dashed line) which gives a good fit to the $F(x = 0) / F_{\mathrm{peak}} = 1/2$ value (solid contour). The main important feature is that $f_{\rm c, crit} \approx 5-50$ is \textit{small}.
For instance, \Lya RT among clumps possessing a hydrogen column density of $N_{\cl,\mathrm{H}} \sim 10^{17}\cm^{-2}$ behaves as inside a homogeneous medium\footnote{This threshold essentially holds for any cloud with a significant neutral fraction. For a cloud which is self-shielding to ionizing radiation, $N_{\rm HI} > x_{\rm HI,photo}^{-1} N_{\rm LLS}$, where $x_{\rm HI,photo} \ll 1$ is the HI fraction in the optically thin photosphere, and $N_{\rm HI, LLS} \sim 10^{17} {\rm cm^{-2}}$.} 
if $N_{\HI, \mathrm{total}} > 10^{18}\cm^{-2}$  (as this implies $\fc > \fccrit$).
Thus, for HI column densities typical of galaxies, a fog of cloudlets is indistinguishable from a homogeneous slab, from the point of view of Ly$\alpha$ RT, explaining why this seemingly oversimplified model is often such a good fit to observations.\\

This convergence towards the homogeneous slab for $\fc\gtrsim \fccrit$ is not only true in the static case. Fig.~\ref{fig:outflows} shows several examples of spectra from slabs with outflows. Specifically, we introduced a linear scaling between $0\kms$ at the half-height to $v_{\mathrm{max}}$ at the boundaries of the slab. In each panel of Fig.~\ref{fig:outflows} we show spectra from clumpy media with different values of \fc and compare them to the spectra from a homogeneously filled slab with the same kinematics and total column densities (black dashed lines in Fig.~\ref{fig:outflows}). The spectra for $\fc \sim 1000$ match the homogeneous spectra very well even for large outflow velocities. This result implies that observed spectra which can be reproduced using homogeneous media and a (complex) velocity structure can also be interpreted in the light of many tiny clumps moving with this velocity. The latter interpretation is more reasonable considering the high (super-sonic) velocities often invoked (through unusually large Doppler parameters) when modelling \Lya spectra. These occur more naturally in a `fog' since the clumps are comoving with the hot gas of the medium \citep{McCourt2016},  and would be hard to understand otherwise since cold gas should be destroyed by shear driven instabilities.

\section{Implications}
\label{sec:implications}

Our results show that \Lya spectra emerging from multiphase media with large covering factors match those emerging from homogeneous media. We identified analytically that this transition between the two RT regimes occurs at \fccrit as given by Equation~\eqref{eq:fccrit}.
This result potentially offers natural explanations to the puzzling mismatches between observed \Lya spectra and the ones from simplified or low-resolution hydrodynamic multiphase models discussed in Sec.~\ref{sec:intro}.

While observed \Lya spectra show a variety of characteristics, there are several common features. One example of a common feature is that the vast majority of all spectra are shifted redwards and are asymmetric with a stronger red than blue component \citep{Steidel2010a,Kulas2011,Erb2014,Trainor2015,Hashimoto2015}. There are roughly as many single- as double peaked spectra \citep[e.g., ][]{Kulas2011,Trainor2015}, and observations indicate that the flux in the `valley' between the peaks is small  \citep{Henry2015,Yang2015}.  
In addition, the peaks themselves commonly show an asymmetry with extended wings; another feature which points towards strong, systematic RT effects\footnote{It is, for instance, difficult to explain the low flux at line center and the asymmetry in the spectrum if some photons experience a very low optical depth.}.
A successful model has to be able to reproduce these observations: the consistent asymmetry of the spectral line, and if double-peaked, the low flux in the valley. \\

Models for Ly$\alpha$ transfer through multiphase media have not yet matched all of these constraints.
The hot, ionized phase provides low-$N_{\rm HI}$ channels through which a significant fraction of the flux can escape. This applies both to simplified clumpy models with low \fc \citep{Hansen2005,2013ApJ...766..124L,Gronke2014a,Gronke2016a}, and to snapshots from hydrodynamical simulations \citep[e.g.,][]{Laursen2009,Zheng2009,Verhamme2012,Behrens2014a,Trebitsch2016}, which typically have $\gg 1 $pc resolution -- not enough to resolve the shattering process \citep[][]{McCourt2016}. As a consequence, hydrodynamical simulations likely give a too low a value for \fc for any given total HI column density, which then generally gives rise to Ly$\alpha$ spectra which possess a too large flux at line center and / or do not show sufficient asymmetries.\\

Our results hint that {\it if} astrophysical systems do indeed possess similar small-scale structures, that these affect \Lya photons in a way that more closely resembles a homogeneous medium. This may provide some new insights into the success of the `shell-model', though this needs to made more quantitative. An additional advantage of this interpretation of this shell-model, is that it may alleviate potential problems with the values inferred by `shell-model fitting'. Examples include unnaturally large Doppler parameters $b$, and/or intrinsic spectra which are wider than their H$\alpha$ counterparts \citep[e.g.,][]{Hashimoto2015,Yang2015}. 
The former (as well as highly supersonic expansion) can be explained if the cold clumps are comoving with the hot gas, and the latter
could hint at RT effects prior to scattering through the multiphase outflow. We will explore complications such as bulk flows, turbulence, and dust in more detail in future work.

\acknowledgments
This research made use of a number of open source software \citep{PER-GRA:2007,Hunter:2007,jones_scipy_2001}. MM and SPO acknowledge NASA grant NNX15AK81G. MM was partially supported by NASA grant HST-HF2-51376.001-A, under NASA contract NAS5-26555. 

\end{document}